\documentclass[final,3p,times]{elsarticle}

\usepackage{amsmath}                   
\usepackage{amssymb}
\usepackage{graphicx}                  
\usepackage{epstopdf}                  
\usepackage{dcolumn}                   
\usepackage{multirow}
\usepackage{rotating}
\usepackage{lineno,hyperref}

\begin{document}

\begin{frontmatter}

\title{Correlated structure of nuclear symmetry energy from covariant nucleon self-energy}

\author[mymainaddress,mysecondaryaddress]{Zhi Wei Liu}

\author[mymainaddress,mysecondaryaddress]{Qian Zhao}

\author[mymainaddress,mysecondaryaddress]{Bao Yuan Sun \corref{mycorrespondingauthor}}
\cortext[mycorrespondingauthor]{Corresponding author}
\ead{sunby@lzu.edu.cn}

\address[mymainaddress]{School of Nuclear Science and Technology, Lanzhou University, 730000 Lanzhou, China}
\address[mysecondaryaddress]{Engineering Research Center for Neutron Application Technology, Ministry of Education, Lanzhou University, Lanzhou 730000, China}

\begin{abstract}
Based on the Hugenholtz-Van Hove theorem, the symmetry energy $J$ and its density slope parameter $L$ are decomposed in terms of the nucleon self-energies within the covariant density functional (CDF) theory. It is found that two structural connections between the different ingredients of $J$ and $L$ construct the fundamental correlation between $L$ and $J$ in the relativistic covariant framework, while the additional contribution from the isovector scalar channel of nucleon-nucleon interaction and those from the second-order symmetry self-energies lead to a deviation, especially the latter limits severely its correlation coefficient and confidence level. In addition, the relationship between the Landau mass $M_L^*$ and the Dirac mass $M_D^*$ is approximated to a reliable linear correlation, which is demonstrate to be sensitive to the momentum dependence of the nucleon self-energies.
\end{abstract}


\end{frontmatter}

\section{Introduction}
The nuclear symmetry energy, which is intrinsically related to the isospin dependence of in-medium nuclear interaction, is not only a fundamental quantity in nuclear physics but also in astrophysics \cite{Lattimer2000PR333.121, Baran2005PR410.335, Steiner2005PR411.325, Li2008PR464.113}. Although significant progress has been made both experimentally and theoretically, our current knowledge about the symmetry energy is still rather poor, especially its density dependence. For instance, a recent data collective analysis constrained the most probable values of symmetry energy at the saturation density as $J=31.7\pm3.2$ MeV and its density slope parameter as $L=58.7\pm28.1$ MeV, retaining a relatively large error bar \cite{Oertel2017RMP89.015007}, implying a big challenge in the community. In the past years, the relationship between symmetry energy $J$ and its density slope parameter $L$ at the saturation density has attracted wide interest in theory. Based on Skyrme and relativistic mean-field models \cite{Farine1978NPA304.317, Ducoin2011PRC83.045810, Dutra2012PRC85.035201, Dutra2014PRC90.055203}, random phase approximation models \cite{Carbone2010PRC81.041301}, quantum Monte Carlo method with realistic interactions \cite{Gandolfi2012PRC85.032801} and other approaches \cite{Oyamatsu2003NPA718.363, Lattimer2013AJ771.51, Horowitz2014JPG41.093001, Blaschke2016arXiv1604.08575, Tews2017AJ848.105}, a possible correlation between $L$ and $J$ has been revealed, but this correlation is relatively weak and its physical mechanism is still unclear. Furthermore, although analytical expressions for $L$ as a linear function of $J$ have been suggested in some special cases \cite{Santos2014PRC90.035203, Mondal2017PRC96.021302, Holt2018PLB784.77}, a more general and model independent linear correlation has not been verified yet. It is thus of necessity to study the universal and more accurate knowledge about this correlation, which enables us to learn more about $L$ with the relatively strict constraint of $J$.

In recent years, based on the Hugenholtz-Van Hove (HVH) theorem \cite{Hugenholtz1958Phys24.363, Satpathy1999PR319.85}, the nuclear symmetry energy and its density slope parameter at arbitrary density have been analytically decomposed in terms of the single-nucleon potential in asymmetric nuclear matter within non-relativistic framework \cite{Xu2011NPA865.1, Chen2012PRC85.024305}, which can be constrained further by the nucleon global optical model potential extracted from nucleon-nucleus and $(p,n)$ charge-exchange reactions \cite{Xu2010PRC82.054607, Li2013PLB721.101}. In addition, the relativistic version of the decomposition of the nuclear symmetry energy and its density slope parameter on the basis of Lorentz covariant nucleon self-energies has been also obtained, and the preliminary application has been performed within nonlinear relativistic mean-field model \cite{Cai2012PLB711.104, Li2018PPNP99.29}. It should point out that since the HVH theorem is valid for any interacting self-bound infinite Fermi system \cite{Hugenholtz1958Phys24.363, Satpathy1999PR319.85}, these decompositions are independent of the detailed nature of the nucleon-nucleon interactions. Therefore, the decomposition of symmetry energy in terms of the nucleon self-energies provide a physically more transparent approach to help us better understand the microscopic origins of the nuclear symmetry energy and its density slope parameter as well as the correlation between them.

The covariant density functional (CDF) theory is great successful in the description of nuclear many-body problem as a relativistic energy density functional (EDF) \cite{Reinhard1989RPP52.439, Ring1996PPNP37.193, Vretenar2005PR409.101, Meng2006PPNP57.470, Liang2015PR570.1, Meng2016}. For example, two important ingredients of nuclear force --- the spin-orbit coupling and the tensor interaction can be both included naturally \cite{Jiang2015PRC91.034326, Zong2018CPC42.24101}. In CDF theory, the nucleon self-energy, which is obtained by the variation of the potential part of EDF with respect to Dirac spinor, can be used to describe the in-medium nuclear interaction effectively. According to this spirit, the nucleon self-energy is dependent on the detailed nature of various versions of CDF due to the uncertainty of in-medium nuclear interaction. For example, (1) besides including nonlinear coupling channels in the Lagrangian \cite{Boguta1977NPA292.413, Boguta1983PLB120.289}, the medium dependence of nucleon self-energy can also be introduced by assuming explicit density dependence of interaction vertices \cite{Typel1999NPA656.331, Long2004PRC69.034319}. (2) Due to the lack of sufficient experimental information, whether the isovector scalar channel of nucleon self-energy should be considered in the CDF theory is still an open question \cite{Roca2011PRC84.054309}. (3) In general, the momentum dependence (non-locality effects) of the nucleon self-energy vanish automatically in the Hartree mean-field level, but inclusion of the Fock diagrams in the CDF theory is a self-consistent way to preserve this dependence \cite{Long2006PLB640.150}. In fact, the uncertainty of the nucleon self-energy in CDF theory brings an opportunity to the future investigation of model independence, such as the universal correlation between
$L$ and $J$.

Based on different commonly used parameter sets in the different versions of the CDF, the symmetry energy $J$ and its density slope parameter $L$ at the saturation density will be decomposed in terms of the Lorentz covariant nucleon self-energies in this paper. Then the possible correlated structure of nuclear symmetry energy and its physical mechanism will be revealed and analysed within relativistic framework, which is the main motivation of the present work.

\section{Theoretical Framework}
Because of the translational and rotational invariance in the rest frame of infinite nuclear matter, and the assumed invariance under parity and time reversal, as well as hermiticity, the nucleon self-energy in asymmetric nuclear matter may be written generally as \cite{Serot1986ANP16.1, Horowitz1983NPA399.529, Horowitz1987NPA464.613, Jaminon1981NPA365.371}
\begin{align}\label{eq:self-energy}
  \Sigma^\tau(\rho,\delta,|\boldsymbol{k}|)&=\Sigma_S^\tau(\rho,\delta,|\boldsymbol{k}|)+\gamma^0\Sigma_0^\tau(\rho,\delta,|\boldsymbol{k}|)\nonumber\\
  &+\boldsymbol{\gamma}\cdot\boldsymbol{\hat{k}}\Sigma_V^\tau(\rho,\delta,|\boldsymbol{k}|),
\end{align}
where isospin index $\tau=1~(-1)$ denotes neutron (proton), $\Sigma_S^\tau$ is the scalar self-energy, $\Sigma_0^\tau$ and $\Sigma_V^\tau$ are the time and space components of the vector self-energy respectively, and $\boldsymbol{\hat{k}}$ is the unit vector along $\boldsymbol{k}$. In general, self-energies depend on the baryon density $\rho=\rho_n+\rho_p$ with $\rho_n$ and $\rho_p$ corresponding to the neutron and proton densities respectively, the isospin asymmetry $\delta=(\rho_n-\rho_p)/\rho$, and the magnitude of the nucleon momentum $|\boldsymbol{k}|$. Besides, self-energies can be divided further according to the contributions from Hartree (direct) and Fock (exchange) diagrams \cite{Sun2008PRC78.065805}, namely,
\begin{subequations}\label{eq:HF}
  \begin{align}
    \Sigma_S^\tau(\rho,\delta,|\boldsymbol{k}|)&=\Sigma_S^{\tau,D}(\rho,\delta)+\Sigma_S^{\tau,E}(\rho,\delta,|\boldsymbol{k}|),\\
    \Sigma_0^\tau(\rho,\delta,|\boldsymbol{k}|)&=\Sigma_0^{\tau,D}(\rho,\delta)+\Sigma_0^{\tau,E}(\rho,\delta,|\boldsymbol{k}|),\\
    \Sigma_V^\tau(\rho,\delta,|\boldsymbol{k}|)&=\Sigma_V^{\tau,E}(\rho,\delta,|\boldsymbol{k}|),
  \end{align}
\end{subequations}
where $\Sigma_\mathcal{O}^{\tau,D}$ and $\Sigma_\mathcal{O}^{\tau,E}$ ($\mathcal{O}$ denotes $S$, $0$ or $V$.) correspond to the Hartree and Fock terms of self-energies. With the density-dependent interaction vertices, the additional rearrangement term $\Sigma_R^\tau(\rho,\delta)$ has to be taken into account properly \cite{Typel1999NPA656.331}. In order to discuss the following results conveniently, the first (second) order symmetry self-energy $\Sigma_\mathcal{O}^{{\rm sym},1}$ ($\Sigma_\mathcal{O}^{{\rm sym},2}$) is defined as
\begin{align}\label{eq:symmetry self-energy}
  \Sigma_\mathcal{O}^{{\rm sym},1}\equiv\frac{1}{2}\left.\frac{\partial\Sigma_{\mathcal{O}}^{-}}{\partial\delta}\right|_{\delta=0},~~~
  \Sigma_\mathcal{O}^{{\rm sym},2}\equiv\frac{1}{4}\left.\frac{\partial^2\Sigma_{\mathcal{O}}^{+}}{\partial\delta^2}\right|_{\delta=0},
\end{align}
where $\Sigma_{\mathcal{O}}^{-}=\Sigma_{\mathcal{O}}^{n}-\Sigma_{\mathcal{O}}^{p}$ and $\Sigma_{\mathcal{O}}^{+}=\Sigma_{\mathcal{O}}^{n}+\Sigma_{\mathcal{O}}^{p}$ correspond to the splitting and sum of neutron-proton self-energies respectively.

With the general form of self-energy, the single-nucleon Dirac equation in nuclear matter can be written as
\begin{align}
  (\boldsymbol{\gamma}\cdot\boldsymbol{k}^{*,\tau}+M^{*,\tau}_D)u(k,s,\tau)=\gamma_0\varepsilon^{*,\tau}u(k,s,\tau),
\end{align}
with the starred quantities
\begin{subequations}
  \begin{align}
    \boldsymbol{k}^{*,\tau}&~=~\boldsymbol{k}~~+~\boldsymbol{\hat{k}}\Sigma_V^\tau(\rho,\delta,|\boldsymbol{k}|),\\
    M^{*,\tau}_D&~=~M~+~~\Sigma_S^\tau(\rho,\delta,|\boldsymbol{k}|),\\
    \varepsilon^{*,\tau}&~=~\varepsilon^\tau~-~~\Sigma_0^\tau(\rho,\delta,|\boldsymbol{k}|),
  \end{align}
\end{subequations}
which satisfy the effective relativistic mass-energy relation $(\varepsilon^{*,\tau})^2=(\boldsymbol{k}^{*,\tau})^2+(M^{*,\tau}_D)^2$. It should be also emphasized that $M^{*,\tau}_D$ is the Dirac mass ( also called the Lorentz-scalar effective mass ). Accordingly, the relativistic single-nucleon energy $\varepsilon^\tau$ can be expressed as
\begin{align}\label{eq:single-nucleon energy}
  \varepsilon^\tau(\rho,\delta,|\boldsymbol{k}|)&=\sqrt{[M^{*,\tau}_D(\rho,\delta,|\boldsymbol{k}|)]^2+[\boldsymbol{k}^{*,\tau}(\rho,\delta,|\boldsymbol{k}|)]^2}\nonumber\\
  &+\Sigma_0^\tau(\rho,\delta,|\boldsymbol{k}|).
\end{align}

The equation of state (EoS) of asymmetric nuclear matter at zero temperature is defined by its binding energy per nucleon $E(\rho,\delta)$. Conventionally, due to the difficulty of analytical extraction, the nuclear symmetry energy can be approximately extracted by expanding the zero-temperature EoS in a Taylor series with respect to the $\delta$. The convergence of such an isospin-asymmetry expansion has been acknowledged in the self-consistent mean-field calculations, such as the CDF theory (at the first Hartree-Fock level) adopted in this work, but broken overall when second-order perturbative contributions are involved in many-body theory \cite{Wellenhofer2016PRC93.055802}. Within this approximation, the EoS is then expressed as,
\begin{subequations}
  \begin{align}
    E(\rho,\delta)&=E_0(\rho)+E_{\rm{sym}}(\rho)\delta^2+\mathcal{O}(\delta^4),\label{eq:EoS}\\
    E_{\rm{sym}}(\rho)&\equiv\left.\frac{1}{2}\frac{\partial^2E(\rho,\delta)}{\partial\delta^2}\right|_{\delta=0},
  \end{align}
\end{subequations}
where $E_0(\rho)=E(\rho,\delta=0)$ denotes the EoS of symmetric nuclear matter, and the second-order coefficient $E_{\rm{sym}}(\rho)$ is usually defined as the density-dependent symmetry energy. The odd-order terms of the expansion are discarded in Eq. \eqref{eq:EoS} owing to assuming the charge-independence of nuclear force and neglecting the Coulomb interaction in the infinite nuclear matter. Furthermore, around the nuclear saturation density $\rho_0$, the symmetry energy can be expanded as
\begin{subequations}
  \begin{align}
    E_{\rm{sym}}(\rho)&=J+L\chi+\mathcal{O}(\chi^2),\\
    L&=Y(\rho_0),~~~Y(\rho)\equiv\left[3\rho\frac{{\rm d}E_{\rm{sym}}(\rho)}{{\rm d}\rho}\right],
  \end{align}
\end{subequations}
where $\chi=(\rho-\rho_0)/3\rho_0$ is a dimensionless variable, $Y(\rho)$ is the density slope parameter of $E_{\rm{sym}}(\rho)$ at an arbitrary density $\rho$, $J$ and $L$ correspond to the symmetry energy and its density slope parameter at saturation density $\rho_0$.

According to the HVH theorem \cite{Hugenholtz1958Phys24.363, Satpathy1999PR319.85}, the nucleon chemical potential in asymmetric nuclear matter should be equal to its Fermi energy (the single-nucleon energy at Fermi surface), i.e.,
\begin{align}\label{eq:HVH}
  \frac{\partial[\rho E(\rho,\delta)]}{\partial\rho_\tau}+M=\varepsilon_F^\tau(\rho,\delta,k_F^\tau),
\end{align}
where $\varepsilon_F^\tau=\varepsilon^\tau(\rho,\delta,|\boldsymbol{k}|=k_F^\tau)$ is the single-nucleon energy, and $k_F^\tau=k_F(1+\tau\delta)^{1/3}$ is the nucleon Fermi momentum with $k_F=(3\pi^2\rho/2)^{1/3}$ being the Fermi momentum in symmetric nuclear matter at density $\rho$. Substituting Eq. \eqref{eq:EoS} into the left-hand side of Eq. \eqref{eq:HVH}, expanding $\varepsilon_F^\tau$ as a power series in $\delta$ on the right-hand side of Eq. \eqref{eq:HVH}, and then comparing the coefficients of first- and second-order $\delta$ terms on both sides could lead to \cite{Cai2012PLB711.104}
\begin{subequations}
  \begin{align}
    E_{\rm{sym}}(\rho)&=\frac{1}{4}\frac{\rm{d}}{\rm{d}\delta}\Big[\sum_\tau\tau\varepsilon_F^\tau(\rho,\delta,k_F^\tau)\Big]\Big|_{\delta=0},\label{eq:Esym}\\
    Y(\rho)&=\frac{3}{4}\frac{\rm{d}^2}{\rm{d}\delta^2}\Big[\sum_\tau\varepsilon_F^\tau(\rho,\delta,k_F^\tau)\Big]\Big|_{\delta=0}+3E_{\rm{sym}}(\rho).\label{eq:L}
  \end{align}
\end{subequations}
Combining Eq. \eqref{eq:single-nucleon energy} and \eqref{eq:Esym}, the symmetry energy can be decomposed according to the separation of nucleon self-energy, namely,
\begin{subequations}
  \begin{align}
    E_{\rm{sym}}(\rho)&=E_{\rm{sym}}^{\rm{kin}}(\rho)+E_{\rm{sym}}^{\rm{mom}}(\rho)+E_{\rm{sym}}^{\rm{1st}}(\rho)\label{eq:Esym-HVH}\\
    &=\frac{k_F^2}{6M_L^*(\rho)}+E_{\rm{sym}}^{\rm{1st}}(\rho),
  \end{align}
\end{subequations}
where $E_{\rm{sym}}^{\rm{kin}}~(J^{\rm{kin}})$ represents the kinetic contribution, $E_{\rm{sym}}^{\rm{mom}}~(J^{\rm{mom}})$ denotes the contribution from the momentum dependence of the nucleon self-energies, $E_{\rm{sym}}^{\rm{1st}}~(J^{\rm{1st}})$ corresponds to the contribution from the first-order symmetry self-energies (at saturation density $\rho_0$), and $M_L^*\equiv k_F\big[\frac{d|\boldsymbol{k}|}{d\varepsilon^\tau}\big]_{|\boldsymbol{k}|=k_F}$ is the nucleon Landau mass in symmetric nuclear matter \cite{Jaminon1989PRC40.354} which can be expressed as
\begin{align}\label{eq:M-Landau}
  M_L^*(\rho)=k_F\left[\frac{k_F^*}{\varepsilon_F^*}+\frac{M_F^*}{\varepsilon_F^*}\frac{\partial\Sigma_S}{\partial|\boldsymbol{k}|}+\frac{\partial\Sigma_0}{\partial|\boldsymbol{k}|}+\frac{k_F^*}{\varepsilon_F^*}\frac{\partial\Sigma_V}{\partial|\boldsymbol{k}|}\right]_{|\boldsymbol{k}|=k_F}^{-1},
\end{align}
with $\Sigma_\mathcal{O}=\Sigma_\mathcal{O}^\tau(\rho,\delta=0,|\boldsymbol{k}|)$, $k_F^*=k^{*,\tau}(\rho,\delta=0,|\boldsymbol{k}|=k_F)$, $M_F^*=M_D^{*,\tau}(\rho,\delta=0,|\boldsymbol{k}|=k_F)$ and $\varepsilon_F^*=\varepsilon^{*,\tau}(\rho,\delta=0,|\boldsymbol{k}|=k_F)$. Similarly, by substituting Eq. \eqref{eq:single-nucleon energy} into Eq. \eqref{eq:L}, the density slope parameter can be also decomposed as
\begin{align}\label{eq:L-HVH}
  Y(\rho)&=Y^{\rm{kin}}(\rho)+Y^{\rm{mom}}(\rho)\nonumber\\
  &+Y^{\rm{1st}}(\rho)+Y^{\rm{cross}}(\rho)+Y^{\rm{2nd}}(\rho),
\end{align}
where $Y^{\rm{kin}}~(L^{\rm{kin}})$ is the kinetic contribution, $Y^{\rm{mom}}~(L^{\rm{mom}})$ denotes the contribution from the momentum dependence of the nucleon self-energies, and $Y^{\rm{1st}}~(L^{\rm{1st}})$ and $Y^{\rm{2nd}}~(L^{\rm{2nd}})$ correspond to the contributions from the first- and second-order symmetry self-energies respectively, while the cross-term $Y^{\rm{cross}}~(L^{\rm{cross}})$ comes from the momentum dependence of the first-order symmetry self-energies (at saturation density $\rho_0$). Their complex analytical expressions are neglected in this paper, and one could refer to Ref. \cite{Cai2012PLB711.104} for more details.

\section{Results and Discussion}
For the different versions of the CDF considered in this paper, we mainly consider parameter sets commonly and successfully used in the description of nuclear matter and finite nuclei. In particular, we select the parameter sets PKA1, PKO1, PKO2, PKO3 for the relativistic Hartree-Fock (RHF) model; TW99, PKDD, DD-ME1, DD-ME2, DD-ME$\delta$, DDH$\delta$ for the density-dependent relativistic mean-field (DD-RMF) model; FSUGold, IU-FSU, TM1, TM2, PK1, NL1, NL3, NLSH, NLZ, NLZ2, NL$\rho$, NL$\rho\delta$, HA for the nonlinear relativistic mean-field (NL-RMF) model; PC-PK1, PC-F1, PC-F2, PC-F3, PC-F4, PC-LA and DD-PC1, FKVW for the nonlinear and density-dependent point-coupling relativistic mean-field (PC-RMF) models, respectively \cite{Li2008PR464.113, Dutra2014PRC90.055203, Zhao2010PRC82.054319, Liu2018PRC97.025801}. Notice that, due to the limitation of the approach itself, the $\pi$ and $\rho$-tensor couplings or the corresponding interaction vertices are missing in all selected RMF functionals, while the RHF ones PKO1, PKO2, and PKO3 contain the $\pi$ couplings, and PKA1 contains both. Besides, all selected 31 parameter sets include the isovector vector channel involving either the isovector vector $\rho$ meson or the isovector vector interaction vertices in the Lagrangian. While the parameter sets DD-ME$\delta$, DDH$\delta$, NL$\rho\delta$, HA further include the isovector scalar meson field $\delta$, and PC-F2, PC-F4, PC-LA and FKVW include the isovector scalar interaction vertices in the Lagrangian.

\subsection{Symmetry energy and its density slope parameter}
\begin{figure}[h]
  \centering
  \includegraphics[width=0.48\textwidth]{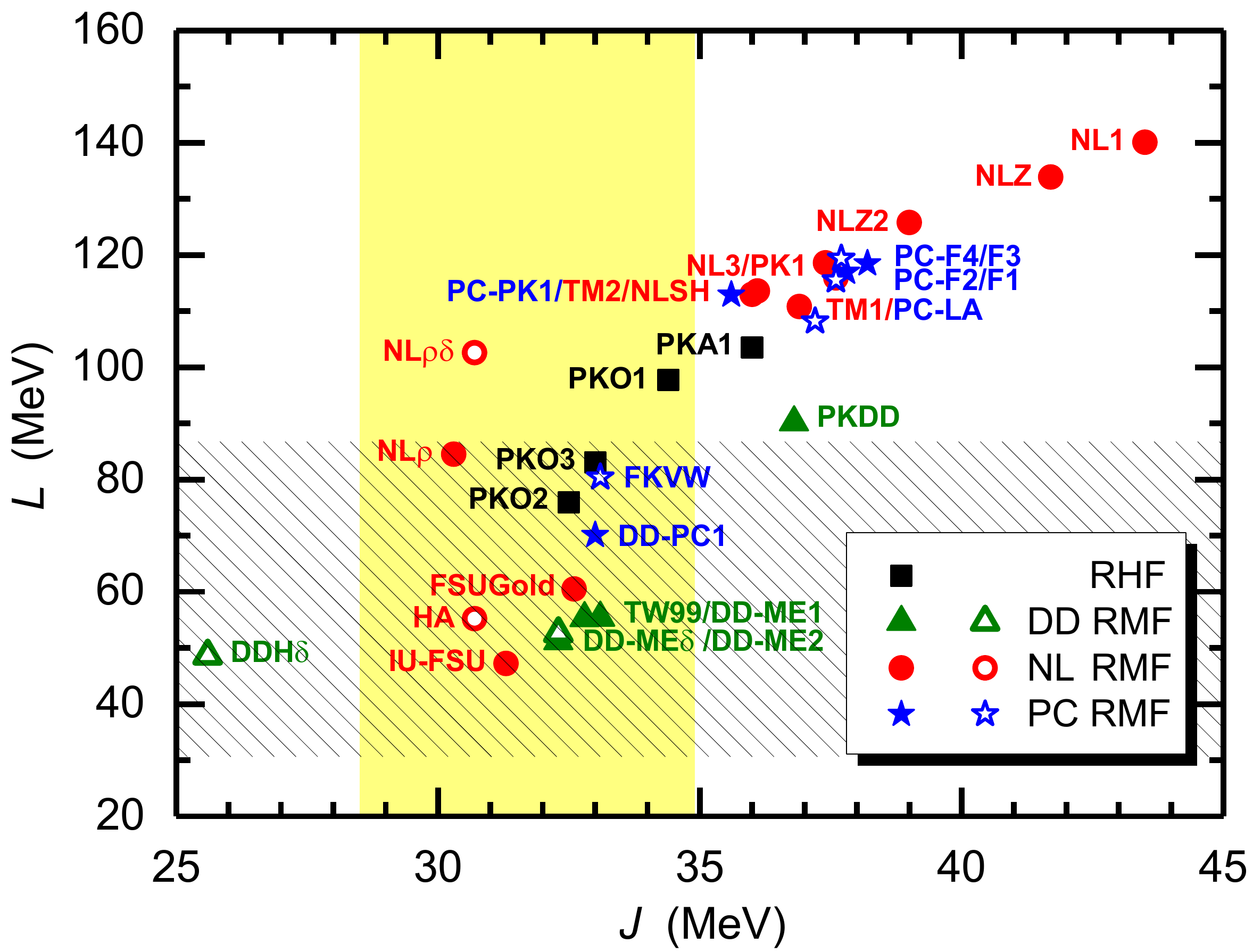}
  \caption{(Color online). Density slope parameter $L$ of symmetry energy against the symmetry energy $J$. The results are calculated with RHF (black squares), DD-RMF (green triangles), NL-RMF (red circles) and PC-RMF (blue stars) models, and the empty points are from the corresponding models with the inclusion of the isovector scalar channel of nucleon-nucleon interaction. The yellow (gray) and shadowed regions depict constraints on $J=31.7\pm3.2$ MeV and $L=58.7\pm28.1$ MeV from Ref. \cite{Oertel2017RMP89.015007}, respectively.
  }\label{Fig:L-J}
\end{figure}

It is worthwhile to review the bulk properties of symmetric nuclear matter at saturation density with all selected CDFs. The saturation density $\rho_0$ and the binding energy per particle $E_0$ are around 0.151 $\rm{fm}^{-3}$ and -16.2 MeV respectively. Utilizing the weighted average method, the symmetry energy $J$, Dirac mass $M_D^*/M$ and Landau mass $M_L^*/M$ are estimated to be about $33.74^{+9.76}_{-8.14}$ MeV, $0.61^{+0.14}_{-0.06}$ and $0.67^{+0.13}_{-0.05}$ respectively. However, the density slope parameter of symmetry energy $L$ lies in a very wide range $85.31^{+54.79}_{-38.11}$ MeV, which leads to a large uncertainty of the information on the symmetry energy at both high and low densities. Figure \ref{Fig:L-J} shows the symmetry energy $J$ versus its density slope parameter $L$ with various CDFs, a recent constraint extracted from the data collective analysis is also plotted \cite{Oertel2017RMP89.015007}. It is seen that although a part of results violate the constraint, there is still a possible correlation between $L$ and $J$ within relativistic framework. As expected, this correlation is fairly weak due to the obvious model dependence of the nucleon self-energies among the selected CDFs.

\begin{figure}[h]
  \centering
  \includegraphics[width=0.48\textwidth]{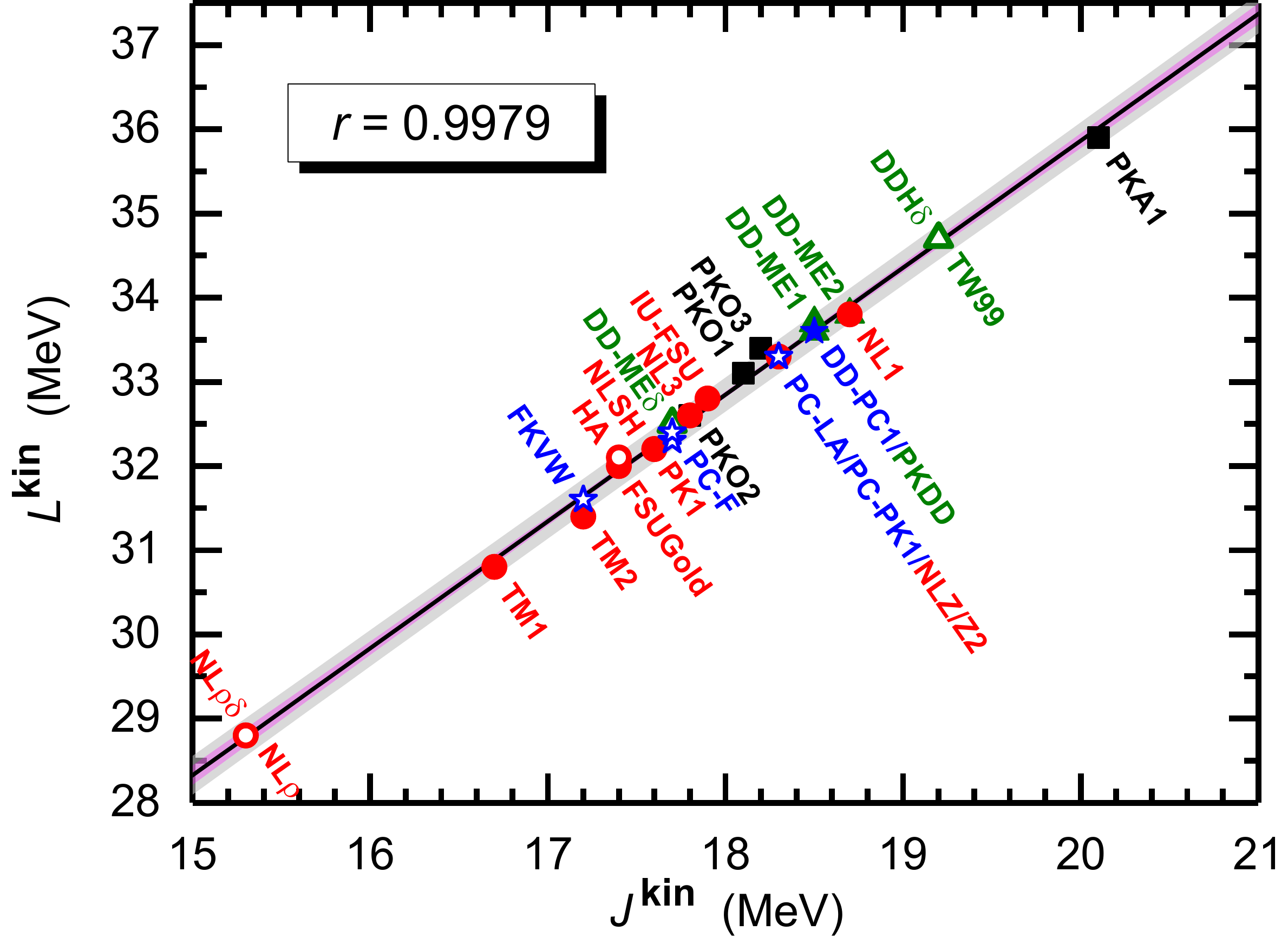}
  \caption{(Color online). Correlation between component $L^{\rm{kin}}$ of the density slope parameter $L$ and component $J^{\rm{kin}}$ of the symmetry energy $J$. The linear fit $L^{\rm{kin}}=1.51J^{\rm{kin}}+5.72$ MeV is obtained with all selected functionals. The inner (outer) colored regions depict the 95\% confidence (prediction) intervals of the linear regression, and Pearson's correlation coefficient \emph{r} is also displayed (see, e.g., Chap. 3 of \cite{Draper1998ARA}).
  }\label{Fig:L-J_details1}
\end{figure}

In order to clarify the physical mechanism of the above correlation, the symmetry energy $J$ and its density slope parameter $L$ are divided into various components according to Eq. \eqref{eq:Esym-HVH} and \eqref{eq:L-HVH} respectively, namely $J=J^{{\rm kin}}+J^{{\rm mom}}+J^{{\rm 1st}}$ and $L=L^{{\rm kin}}+L^{{\rm mom}}+L^{{\rm 1st}}+L^{{\rm cross}}+L^{{\rm 2nd}}$. It is found that there is a strong likelihood that $L^{\rm{kin}}$ calculated with all selected functionals is linearly correlated to the corresponding $J^{\rm{kin}}$, as shown in Figure \ref{Fig:L-J_details1}. In fact, this underlying and reliable linear correlation could be better understood based on the decomposition of symmetry energy in terms of the nucleon self-energies. Utilizing the least-square method, the fitting result $L^{\rm{kin}}=1.51J^{\rm{kin}}+5.72$ MeV, and its Pearson correlation coefficient \emph{r} = 0.9979 indicate that
\begin{align}
  -\frac{1}{2}\frac{k_F}{6}\frac{k_F^*}{\varepsilon_F^*}+\frac{k_F^2}{6\varepsilon_F^*}\frac{M_F^{*2}}{\varepsilon_F^{*2}}\approx 5.72~{\rm MeV}\nonumber
\end{align}
is almost independent of the selected functional. In addition, the relationship between the residual components, namely $(L^{{\rm mom}}+L^{{\rm 1st}}+L^{{\rm cross}}+L^{{\rm 2nd}})$ and $(J^{{\rm mom}}+J^{{\rm 1st}})$ is also investigated, but a strong correlation is not found. Further quantitative analysis shows that $L^{{\rm 2nd}}=2.87^{+29.03}_{-36.17}$ contributes the main part of the uncertainty of $L$, even if its weighted average is quite small, which is qualitatively consistent with the result extracted from neutron-nucleus scattering data \cite{Li2013PLB721.101}.

\begin{figure}[h]
  \centering
  \includegraphics[width=0.48\textwidth]{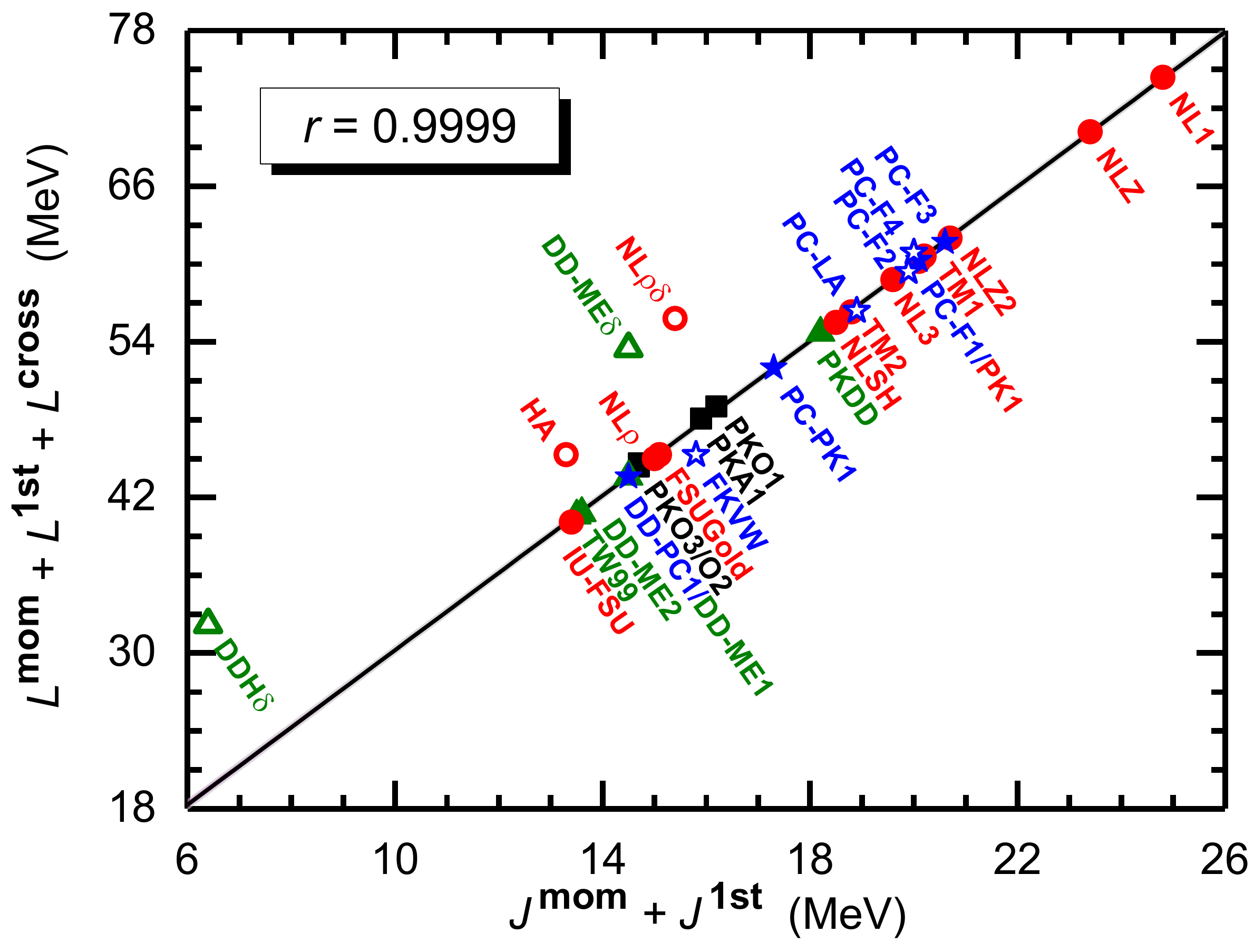}
  \caption{(Color online). Correlation between components $(L^{\rm{mom}}+L^{\rm{1st}}+L^{\rm{cross}})$ of the density slope parameter $L$ and components $(J^{\rm{mom}}+J^{\rm{1st}})$ of the symmetry energy $J$. The linear fit $L^{\rm{mom}}+L^{\rm{1st}}+L^{\rm{cross}}=2.98(J^{\rm{mom}}+J^{\rm{1st}})+0.45$ MeV is obtained with a part of selected functionals, where the models with the isovector scalar channel of nucleon-nucleon interaction are not included.
  }\label{Fig:L-J_details2}
\end{figure}

After neglecting the uncertain $L^{{\rm 2nd}}$, Fig. \ref{Fig:L-J_details2} reveals another underlying and reliable linear correlation between $(L^{\rm{mom}}+L^{\rm{1st}}+L^{\rm{cross}})$ and $(J^{\rm{mom}}+J^{\rm{1st}})$. The fitting result $L^{\rm{mom}}+L^{\rm{1st}}+L^{\rm{cross}}=2.98(J^{\rm{mom}}+J^{\rm{1st}})+0.45~\rm{MeV},\emph{r}=0.9999$ is obtained with a part of selected functionals, where the models with the isovector scalar channel of nucleon-nucleon interaction is not included. Similarly, this linear correlation could also be obtained through theoretical analysis. Owing to the limitation of the Hartree mean-field itself, the space-like self-energy $\Sigma_V$ vanishes while the scalar and time-like self-energies are momentum independent in the general RMF models, thus the above linear correlation could be reduced to a strict proportional function $L^{\rm{1st}}=3J^{\rm{1st}}$. Although the inclusion of the Fock terms in the CDF theory violate strict proportional function, the contribution from the momentum dependence of the nucleon self-energies and the first-order symmetry self-energies weaken this effect.

\begin{figure}[h]
  \centering
  \includegraphics[width=0.48\textwidth]{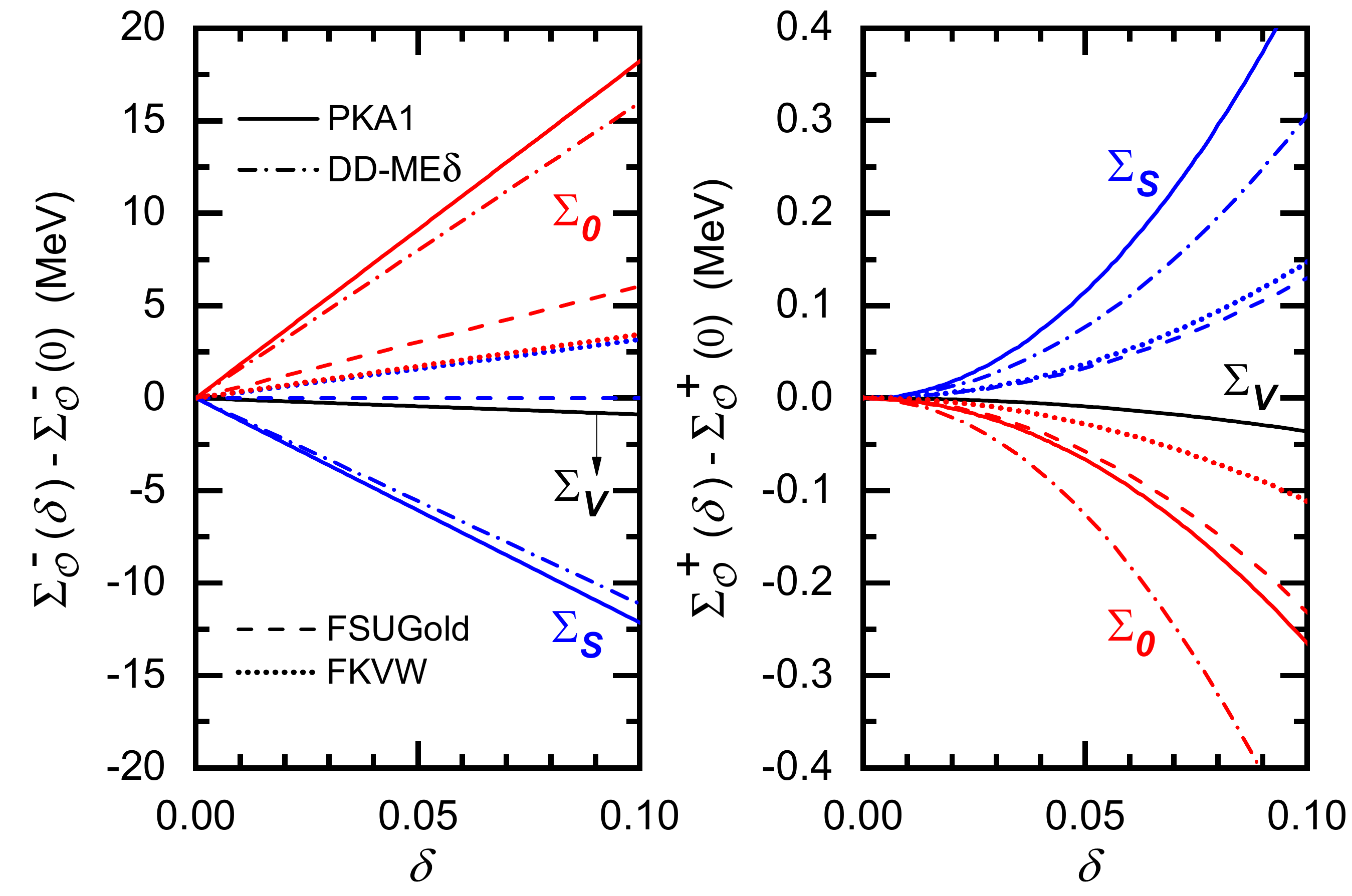}
  \caption{(Color online). (left) Isospin dependence of the splitting of neutron-proton self-energies at saturation density. (right) Isospin dependence of the sum of nucleon self-energies at saturation density.
  }\label{Fig:Sig-iso}
\end{figure}

Shown in Fig. \ref{Fig:Sig-iso} (left panel) are the splitting $\Sigma_\mathcal{O}^-$ as functions of the isospin asymmetry $\delta$ at saturation density, which is related to the first-order symmetry self-energies according to Eq. \eqref{eq:symmetry self-energy}. The inclusion of the isovector scalar channel in the RMF model yields the scalar component of the first-order symmetry self-energy, which leads to the result deviating from the strict proportional function $L^{\rm{1st}}=3J^{\rm{1st}}$. Similarly, shown in Fig. \ref{Fig:Sig-iso} (right panel) are the sum $\Sigma_\mathcal{O}^+$ as functions of the isospin asymmetry $\delta$ at saturation density, which is related to the second-order symmetry self-energies according to Eq. \eqref{eq:symmetry self-energy}. It is seen that the isospin dependence of $\Sigma_\mathcal{O}^+$ is sensitive to in-medium nuclear interaction, namely, $L^{{\rm 2nd}}$ is obviously dependent on the selected functional.

To sum up, the above two underlying and reliable linear correlations construct the fundamental correlation between $L$ and $J$, while the additional contribution from the isovector scalar channel of nucleon-nucleon interaction and the model dependent $L^{\rm{2nd}}$ lead to a deviation, especially the latter limits severely its correlation coefficient and confidence level. Without the inclusion of the isovector scalar channel in the CDF theory, the fitting result $L=2.91J-17.95+L^{\rm{2nd}}~\rm{MeV},\emph{r}=0.9905$ is obtained with other functionals. Moreover, we reanalyze the data in ref. \cite{Chen2012PRC85.024305}, but a reliable correlated structure is not found within non-relativistic framework.

\subsection{Dirac mass and Landau mass}
\begin{figure}[h]
  \centering
  \includegraphics[width=0.48\textwidth]{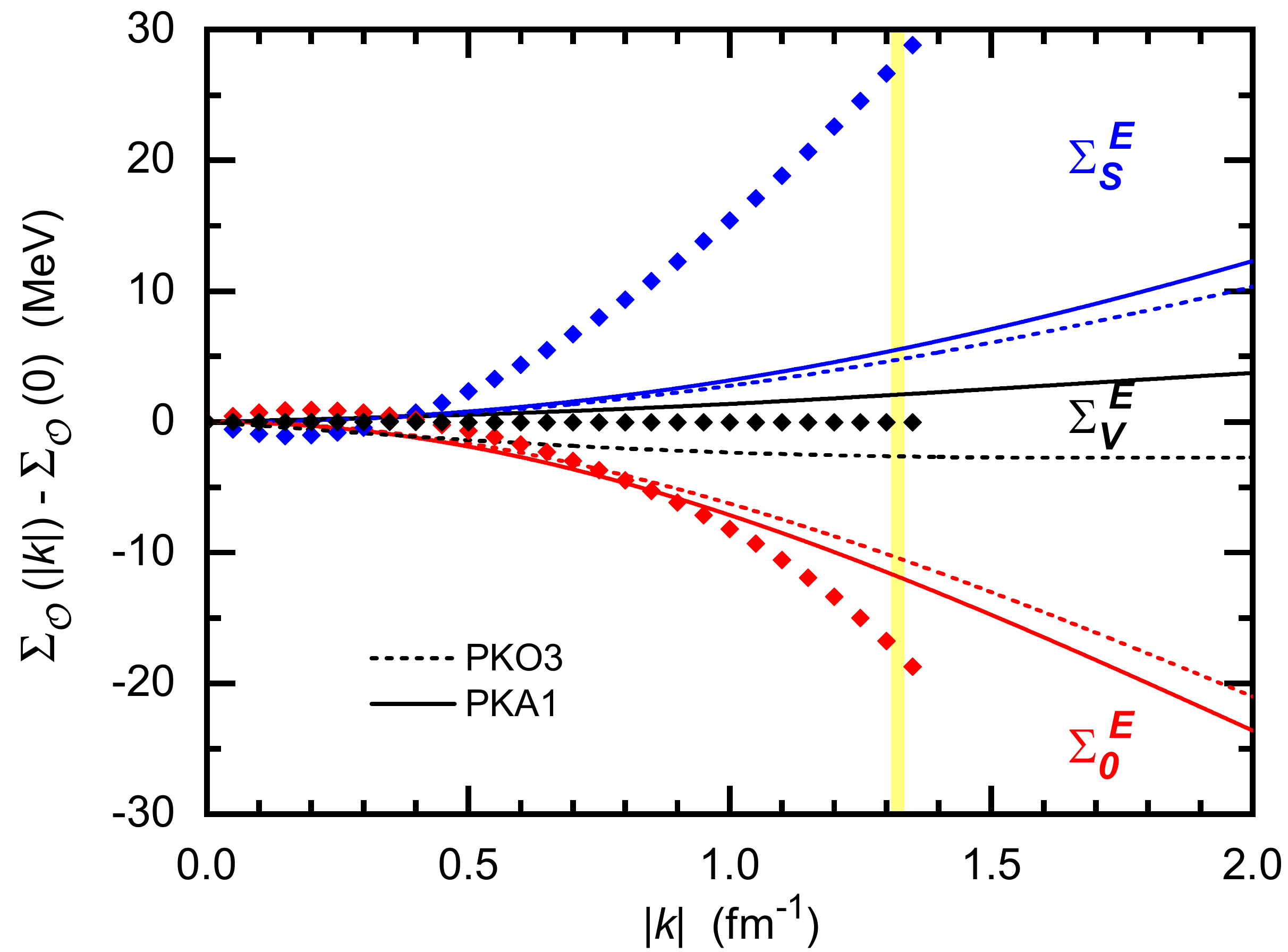}
  \caption{(Color online). Momentum dependence of self-energies from Fock terms at saturation density for symmetric nuclear matter. The results (lines) are calculated with RHF model. Using the parametrizations for self-energies in Dirac-Brueckner-Hartree-Fock model \cite{Katayama2013PRC88.035805}, the corresponding results (diamonds) are plotted in figure for comparison. The yellow (gray) region depicts the corresponding Fermi momentum $1.31\sim1.33~\rm{fm}^{-1}$.
  }\label{Fig:Sig-k}
\end{figure}

As has been discussed, the isospin dependence of the nucleon self-energy is the major factor affecting the correlation between $L$ and $J$, rather than the momentum dependence in the CDF theory. While as an important degree of freedom of the nucleon self-energy, the momentum dependence is expected to affect essentially the Landau mass $M_L^*$, which characterizes the density of states around the Fermi surface. Shown in Fig. \ref{Fig:Sig-k} are the nucleon self-energies calculated with RHF model as functions of the momentum at saturation density for symmetric nuclear matter, compared with the parameterized results within Dirac-Brueckner-Hartree-Fock (DBHF) model \cite{Katayama2013PRC88.035805}. In spite of the momentum dependence of the nucleon self-energies within RHF model is qualitatively consistent with the parameterized results within DBHF model, there are still obvious divergences in quantification, especially for the scalar self-energy.

\begin{figure}[h]
  \centering
  \includegraphics[width=0.48\textwidth]{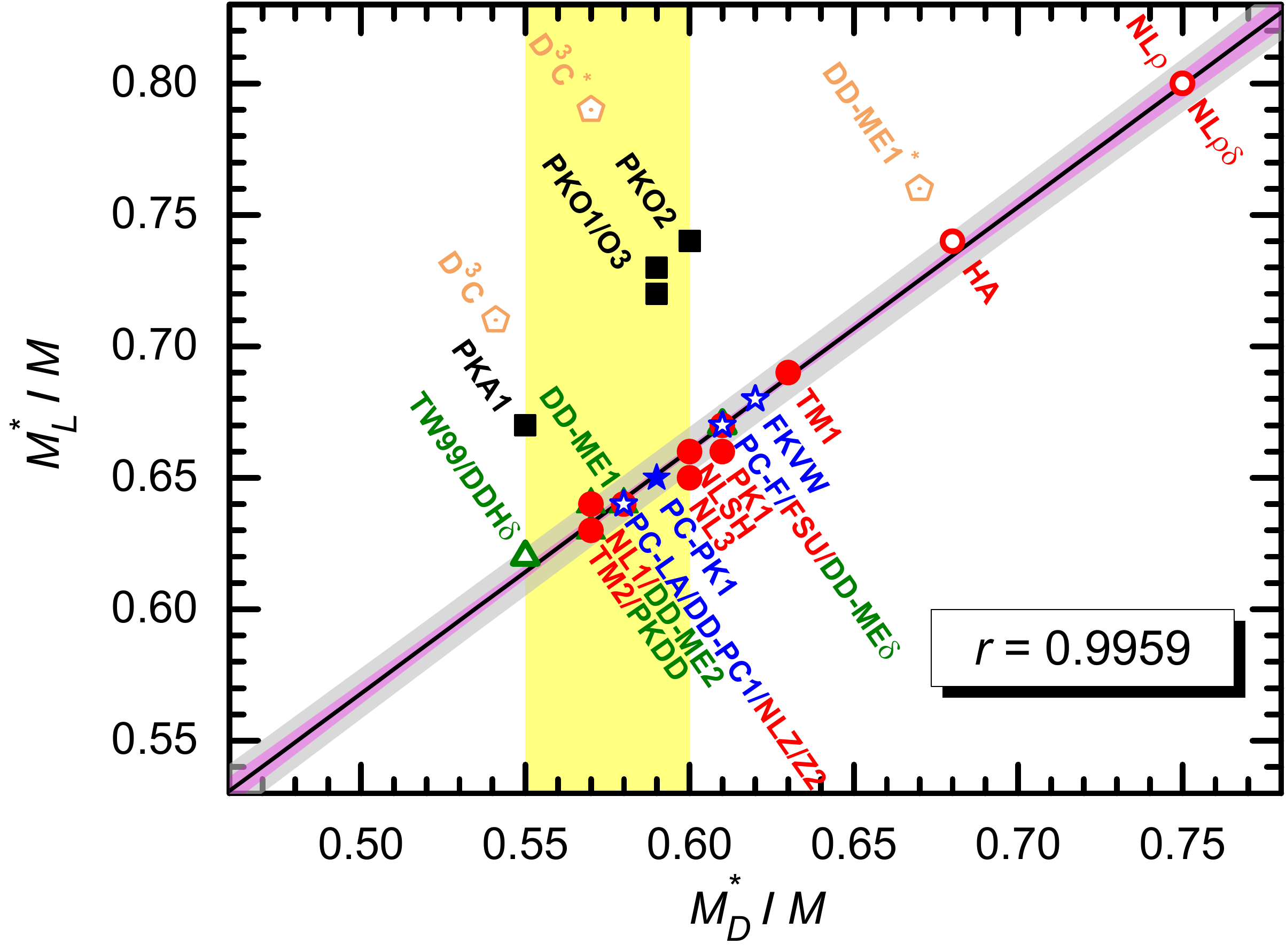}
  \caption{(Color online). Correlation between the Landau mass $M_L^*/M$ and the Dirac mass $M_D^*/M$ at saturation density for symmetric nuclear matter. The linear fit $M_L^*/M=0.93M_D^*/M+0.11$ is obtained with CDFs except RHF model. The referred results DD-ME1$^*$, D$^3$C and D$^3$C$^*$ are from Refs. \cite{Niksic2005PRC71.014308, Typel2005PRC71.064301, Marketin2007PRC75.024304}. The yellow (gray) region depicts the constraint on the Dirac mass $M_D^*/M=0.575\pm0.025$ from the spin-orbit splitting in finite nuclei \cite{Typel2005PRC71.064301, Marketin2007PRC75.024304}.
  }\label{Fig:ML-MD}
\end{figure}

The relation between the Landau mass and the Dirac mass is $M_L^*=\sqrt{k_F^2+M_D^{*2}}$ in the general RMF models, which can be approximated to a linear correlation, as shown in Fig. \ref{Fig:ML-MD}. The fitting result $M_L^*/M=0.93M_D^*/M+0.11,\emph{r}=0.9959$ is obtained with all selected RMF functionals. However, due to the momentum dependence of the nucleon self-energies, especially the stronger dependence of the time component than the scalar one in RHF model, the corresponding results systematically deviate from the above correlation according to Eq. \eqref{eq:M-Landau}.

\section{Summary}
Based on the Hugenholtz-Van Hove theorem, the symmetry energy $J$ and its density slope parameter $L$ are decomposed in terms of the Lorentz covariant nucleon self-energies with various kinds of CDF models. In doing the calculations, one correlated structure of $L^{\rm{kin}}$ with $J^{\rm{kin}}$ and another correlated structure of $(L^{\rm{mom}}+L^{\rm{1st}}+L^{\rm{cross}})$ with $(J^{\rm{mom}}+J^{\rm{1st}})$ appear. These two underlying and reliable linear correlations construct the fundamental correlation between $L$ and $J$ in the relativistic framework, while the additional contribution from the isovector scalar channel of nucleon-nucleon interaction and the model dependent $L^{\rm{2nd}}$ lead to a deviation, especially the latter limits severely its correlation coefficient and confidence level. In addition, the relationship between the Landau mass $M_L^*$ and the Dirac mass $M_D^*$ is approximated to a reliable linear correlation, which is demonstrate to be sensitive to the momentum dependence of the nucleon self-energies.

\section*{Acknowledgements}
This work is partly supported by the National Natural Science Foundation of China (Grant Nos. 11375076 and 11875152) and the Fundamental Research Funds for the Central Universities (Grant No. lzujbky-2016-30).

\section*{References}

\bibliographystyle{elsarticle-num}


\begin{thebibliography}{10}
\expandafter\ifx\csname url\endcsname\relax
  \def\url#1{\texttt{#1}}\fi
\expandafter\ifx\csname urlprefix\endcsname\relax\def\urlprefix{URL }\fi
\expandafter\ifx\csname href\endcsname\relax
  \def\href#1#2{#2} \def\path#1{#1}\fi

\bibitem{Lattimer2000PR333.121}
J.~M. Lattimer, M.~Prakash, Nuclear matter and its role in supernovae, neutron
  stars and compact object binary mergers, Phys. Rep. 333 (2000) 121 -- 146.

\bibitem{Baran2005PR410.335}
V.~Baran, M.~Colonna, V.~Greco, M.~Di~Toro, Reaction dynamics with exotic
  nuclei, Phys. Rep. 410 (2005) 335 -- 466.

\bibitem{Steiner2005PR411.325}
A.~W. Steiner, M.~Prakash, J.~M. Lattimer, P.~J. Ellis, Isospin asymmetry in
  nuclei and neutron stars, Phys. Rep. 411 (2005) 325 -- 375.

\bibitem{Li2008PR464.113}
B.~A. Li, L.~W. Chen, C.~M. Ko, Recent progress and new challenges in isospin
  physics with heavy-ion reactions, Phys. Rep. 464 (2008) 113 -- 281.

\bibitem{Oertel2017RMP89.015007}
M.~Oertel, M.~Hempel, T.~Kl\"ahn, S.~Typel, Equations of state for supernovae
  and compact stars, Rev. Mod. Phys. 89 (2017) 015007.

\bibitem{Farine1978NPA304.317}
M.~Farine, J.~M. Pearson, B.~Rouben, Higher-order volume-symmetry terms of the
  mass formula, Nucl. Phys. A 304 (1978) 317 -- 326.

\bibitem{Ducoin2011PRC83.045810}
C.~Ducoin, J.~Margueron, C.~Provid\^encia, I.~Vida\~na, Core-crust transition
  in neutron stars: Predictivity of density developments, Phys. Rev. C 83
  (2011) 045810.

\bibitem{Dutra2012PRC85.035201}
M.~Dutra, O.~Louren\ifmmode~\mbox{\c{c}}\else \c{c}\fi{}o, J.~S. S\'a~Martins,
  A.~Delfino, J.~R. Stone, P.~D. Stevenson, Skyrme interaction and nuclear
  matter constraints, Phys. Rev. C 85 (2012) 035201.

\bibitem{Dutra2014PRC90.055203}
M.~Dutra, O.~Louren\ifmmode~\mbox{\c{c}}\else \c{c}\fi{}o, S.~S. Avancini,
  B.~V. Carlson, A.~Delfino, D.~P. Menezes, C.~Provid\^encia, S.~Typel, J.~R.
  Stone, Relativistic mean-field hadronic models under nuclear matter
  constraints, Phys. Rev. C 90 (2014) 055203.

\bibitem{Carbone2010PRC81.041301}
A.~Carbone, G.~Col\`o, A.~Bracco, L.~G. Cao, P.~F. Bortignon, F.~Camera,
  O.~Wieland, Constraints on the symmetry energy and neutron skins from pygmy
  resonances in $^{68}\mathrm{Ni}$ and $^{132}\mathrm{Sn}$, Phys. Rev. C 81
  (2010) 041301(R).

\bibitem{Gandolfi2012PRC85.032801}
S.~Gandolfi, J.~Carlson, S.~Reddy, Maximum mass and radius of neutron stars,
  and the nuclear symmetry energy, Phys. Rev. C 85 (2012) 032801(R).

\bibitem{Oyamatsu2003NPA718.363}
K.~Oyamatsu, K.~Iida, Empirical properties of asymmetric nuclear matter to be
  obtained from unstable nuclei, Nucl. Phys. A 718 (2003) 363 -- 366.

\bibitem{Lattimer2013AJ771.51}
J.~M. Lattimer, Y.~Lim, Constraining the symmetry parameters of the nuclear
  interaction, Astrophys. J. 771 (2013) 51.

\bibitem{Horowitz2014JPG41.093001}
C.~J. Horowitz, E.~F. Brown, Y.~Kim, W.~G. Lynch, R.~Michaels, O.~A.,
  J.~Piekarewicz, M.~B. Tsang, H.~H. Wolter, A way forward in the study of the
  symmetry energy: experiment, theory, and observation, J. Phys. G: Nucl. Part.
  Phys. 41 (2014) 093001.

\bibitem{Blaschke2016arXiv1604.08575}
D.~Blaschke, D.~E. Alvarez-Castillo, T.~Klahn, Universal symmetry energy
  contribution to the neutron star equation of state, arXiv 1604 (2016) 08575.

\bibitem{Tews2017AJ848.105}
I.~Tews, J.~M. Lattimer, A.~Ohnishi, E.~E. Kolomeitsev, Symmetry parameter
  constraints from a lower bound on neutron-matter energy, Astrophys. J. 848
  (2017) 105.

\bibitem{Santos2014PRC90.035203}
B.~M. Santos, M.~Dutra, O.~Louren\ifmmode~\mbox{\c{c}}\else \c{c}\fi{}o,
  A.~Delfino, Correlations between the nuclear matter symmetry energy, its
  slope, and curvature from a nonrelativistic solvable approach and beyond,
  Phys. Rev. C 90 (2014) 035203.

\bibitem{Mondal2017PRC96.021302}
C.~Mondal, B.~K. Agrawal, J.~N. De, S.~K. Samaddar, M.~Centelles, X.~Vi\~nas,
  Interdependence of different symmetry energy elements, Phys. Rev. C 96 (2017)
  021302(R).

\bibitem{Holt2018PLB784.77}
J.~W. Holt, Y.~Lim, Universal correlations in the nuclear symmetry energy,
  slope parameter, and curvature, Phys. Lett. B 784 (2018) 77 -- 81.

\bibitem{Hugenholtz1958Phys24.363}
N.~M. Hugenholtz, L.~Van~Hove, A theorem on the single particle energy in a
  fermi gas with interaction, Phys. 24 (1958) 363 -- 376.

\bibitem{Satpathy1999PR319.85}
L.~Satpathy, V.~S. Uma~Maheswari, R.~C. Nayak, Finite nuclei to nuclear matter:
  a leptodermous approach, Phys. Rep. 319 (1999) 85 -- 144.

\bibitem{Xu2011NPA865.1}
C.~Xu, B.~A. Li, L.~W. Chen, C.~M. Ko, Analytical relations between nuclear
  symmetry energy and single-nucleon potentials in isospin asymmetric nuclear
  matter, Nucl. Phys. A 865 (2011) 1 -- 16.

\bibitem{Chen2012PRC85.024305}
R.~Chen, B.~J. Cai, L.~W. Chen, B.~A. Li, X.~H. Li, C.~Xu, Single-nucleon
  potential decomposition of the nuclear symmetry energy, Phys. Rev. C 85
  (2012) 024305.

\bibitem{Xu2010PRC82.054607}
C.~Xu, B.~A. Li, L.~W. Chen, Symmetry energy, its density slope, and
  neutron-proton effective mass splitting at normal density extracted from
  global nucleon optical potentials, Phys. Rev. C 82 (2010) 054607.

\bibitem{Li2013PLB721.101}
X.~H. Li, B.~J. Cai, L.~W. Chen, R.~Chen, B.~A. Li, C.~Xu, Extracting the
  nuclear symmetry potential and energy from neutron-nucleus scattering data,
  Phys. Lett. B 721 (2013) 101 -- 106.

\bibitem{Cai2012PLB711.104}
B.~J. Cai, L.~W. Chen, Lorentz covariant nucleon self-energy decomposition of
  the nuclear symmetry energy, Phys. Lett. B 711 (2012) 104 -- 108.

\bibitem{Li2018PPNP99.29}
B.~A. Li, B.~J. Cai, L.~W. Chen, J.~Xu, Nucleon effective masses in
  neutron-rich matter, Prog. Part. Nucl. Phys 99 (2018) 29 -- 119.

\bibitem{Reinhard1989RPP52.439}
P.~G. Reinhard, The relativistic mean-field description of nuclei and nuclear
  dynamics, Rep. Prog. Phys. 52 (1989) 439.

\bibitem{Ring1996PPNP37.193}
P.~Ring, Relativistic mean field theory in finite nuclei, Prog. Part. Nucl.
  Phys 37 (1996) 193.

\bibitem{Vretenar2005PR409.101}
D.~Vretenar, A.~V. Afanasjev, G.~A. Lalazissis, P.~Ring, Relativistic
  hartree–bogoliubov theory: static and dynamic aspects of exotic nuclear
  structure, Phys. Rep. 409 (2005) 101.

\bibitem{Meng2006PPNP57.470}
J.~Meng, H.~Toki, S.~G. Zhou, S.~Q. Zhang, W.~H. Long, L.~S. Geng, Relativistic
  continuum hartree bogoliubov theory for ground-state properties of exotic
  nuclei, Prog. Part. Nucl. Phys 57 (2006) 470.

\bibitem{Liang2015PR570.1}
H.~Z. Liang, J.~Meng, S.~G. Zhou, Hidden pseudospin and spin symmetries and
  their origins in atomic nuclei, Phys. Rep. 570 (2015) 1.

\bibitem{Meng2016}
J.~Meng, Relativistic Density Functional for Nuclear Structure, World
  Scientific, Singapore, 2016.

\bibitem{Jiang2015PRC91.034326}
L.~J. Jiang, S.~Yang, B.~Y. Sun, W.~H. Long, H.~Q. Gu, Nuclear tensor
  interaction in a covariant energy density functional, Phys. Rev. C 91 (2015)
  034326.

\bibitem{Zong2018CPC42.24101}
Y.~Y. Zong, S.~B. Yuan, Relativistic interpretation of the nature of the
  nuclear tensor force, Chinese physics C 42 (2018) 24101.

\bibitem{Boguta1977NPA292.413}
J.~Boguta, A.~R. Bodmer, Relativistic calculation of nuclear matter and the
  nuclear surface, Nucl. Phys. A 292 (1977) 413 -- 428.

\bibitem{Boguta1983PLB120.289}
J.~Boguta, H.~Stocker, Systematics of nuclear matter properties in a non-linear
  relativistic field theory, Phys. Lett. B 120 (1983) 289 -- 293.

\bibitem{Typel1999NPA656.331}
S.~Typel, H.~H. Wolter, Relativistic mean field calculations with
  density-dependent meson-nucleon coupling, Nucl. Phys. A 656 (1999) 331.

\bibitem{Long2004PRC69.034319}
W.~H. Long, J.~Meng, N.~Van~Giai, S.~G. Zhou, New effective interactions in
  relativistic mean field theory with nonlinear terms and density-dependent
  meson-nucleon coupling, Phys. Rev. C 69 (2004) 034319.

\bibitem{Roca2011PRC84.054309}
X.~Roca-Maza, X.~Vi\~nas, M.~Centelles, P.~Ring, P.~Schuck, Relativistic
  mean-field interaction with density-dependent meson-nucleon vertices based on
  microscopical calculations, Phys. Rev. C 84 (2011) 054309.

\bibitem{Long2006PLB640.150}
W.~H. Long, N.~Van~Giai, J.~Meng, Density-dependent relativistic hartree–fock
  approach, Phys. Lett. B 640 (2006) 150 -- 154.

\bibitem{Serot1986ANP16.1}
B.~D. Serot, J.~D. Walecka, The relativistic nuclear many-body problem, Adv.
  Nucl. Phys. 16 (1986) 1.

\bibitem{Horowitz1983NPA399.529}
C.~J. Horowitz, B.~D. Serot, Properties of nuclear and neutron matter in a
  relativistic hartree-fock theory, Nucl. Phys. A 399 (1983) 529 -- 562.

\bibitem{Horowitz1987NPA464.613}
C.~J. Horowitz, B.~D. Serot, The relativistic two-nucleon problem in nuclear
  matter, Nucl. Phys. A 464 (1987) 613 -- 699.

\bibitem{Jaminon1981NPA365.371}
M.~Jaminon, C.~Mahaux, P.~Rochus, Single-particle potential in a relativistic
  hartree-fock mean field approximation, Nucl. Phys. A 365 (1981) 371 -- 391.

\bibitem{Sun2008PRC78.065805}
B.~Y. Sun, W.~H. Long, J.~Meng, U.~Lombardo, Neutron star properties in
  density-dependent relativistic hartree-fock theory, Phys. Rev. C 78 (2008)
  065805.

\bibitem{Wellenhofer2016PRC93.055802}
C.~Wellenhofer, J.~W. Holt, N.~Kaiser, Divergence of the isospin-asymmetry
  expansion of the nuclear equation of state in many-body perturbation theory,
  Phys. Rev. C 93 (2016) 055802.

\bibitem{Jaminon1989PRC40.354}
M.~Jaminon, C.~Mahaux, Effective masses in relativistic approaches to the
  nucleon-nucleus mean field, Phys. Rev. C 40 (1989) 354--367.

\bibitem{Zhao2010PRC82.054319}
P.~W. Zhao, Z.~P. Li, J.~M. Yao, J.~Meng, New parametrization for the nuclear
  covariant energy density functional with a point-coupling interaction, Phys.
  Rev. C 82 (2010) 054319.

\bibitem{Liu2018PRC97.025801}
Z.~W. Liu, Z.~Qian, R.~Y. Xing, J.~R. Niu, B.~Y. Sun, Nuclear fourth-order
  symmetry energy and its effects on neutron star properties in the
  relativistic hartree-fock theory, Phys. Rev. C 97 (2018) 025801.

\bibitem{Draper1998ARA}
N.~R. Draper, H.~Smith, Applied Regression Analysis, Wiley, New York, 1998.

\bibitem{Katayama2013PRC88.035805}
T.~Katayama, K.~Saito, Properties of dense, asymmetric nuclear matter in
  dirac-brueckner-hartree-fock approach, Phys. Rev. C 88 (2013) 035805.

\bibitem{Niksic2005PRC71.014308}
T.~Nik\ifmmode \check{s}\else \v{s}\fi{}i\ifmmode~\acute{c}\else \'{c}\fi{},
  T.~Marketin, D.~Vretenar, N.~Paar, P.~Ring, \ensuremath{\beta}-decay rates of
  $r$-process nuclei in the relativistic quasiparticle random phase
  approximation, Phys. Rev. C 71 (2005) 014308.

\bibitem{Typel2005PRC71.064301}
S.~Typel, Relativistic model for nuclear matter and atomic nuclei with
  momentum-dependent self-energies, Phys. Rev. C 71 (2005) 064301.

\bibitem{Marketin2007PRC75.024304}
T.~Marketin, D.~Vretenar, P.~Ring, Calculation of \ensuremath{\beta}-decay
  rates in a relativistic model with momentum-dependent self-energies, Phys.
  Rev. C 75 (2007) 024304.

\end{thebibliography}

\end{document}